\documentclass[a4paper,fleqn,usenatbib]{mnras}
\usepackage[T1]{fontenc}
\usepackage{ae,aecompl}
\usepackage{epsf,graphics,graphicx}
\usepackage{amsmath}
\usepackage{amssymb,latexsym,mathrsfs, bm}
\usepackage{color}
\usepackage{float}
\usepackage{times}
\usepackage{pdflscape}
\usepackage{longtable}

\def\eq#1{{Eq.~(\ref{#1})}}

\title[Water masers around binary black holes]{Signatures of supermassive black hole binaries on maser systems}
\author[Padmanabhan and Loeb]{Hamsa
Padmanabhan$^{1}$\thanks{Electronic address: hamsa.padmanabhan@unige.ch} and
Abraham Loeb$^{2}$\thanks{Electronic address: {aloeb@cfa.harvard.edu}
}\\
$^{1}$D\'epartement de Physique Th\'eorique, Universit\'e de Gen\`eve,
24 quai Ernest-Ansermet, CH 1211 Gen\`eve 4, Switzerland\\
$^{2}$Astronomy department, Harvard University,
60 Garden Street, Cambridge, MA 02138, USA
}

\begin{document}
\date{ }
\maketitle

\begin{abstract}
We illustrate a novel signature of black hole binaries from their effect on the kinematics of water maser emission in their environments.  With the help of simulations, we establish the condition for clumps to mase based on their coherence lengths calibrated to those of the known maser galaxy NGC 4258. This is then used to identify masing clumps around a binary black hole system, and quantify the kinematic and spectral differences relative to the single black hole case. For some generic circumstances, blue-shifted masers around a binary black hole are found to preferentially follow the Keplerian rotation curve observed in the single black hole case. The  redshifted ones, however, are found to visibly deviate from this relation, and also display more scatter with a tendency towards lower absolute values of the velocity along the line-of-sight. The spectrum of the masers as a function of line-of-sight velocity also shows a double peaked structure, reminiscent of recent observations of systems such as Mrk 1. Our results motivate future prospects for identifying binary black hole candidates with the help of water maser emissions.
\end{abstract}


\section{Introduction}

Water masers trace dense, warm molecular gas from circumnuclear environments in active galaxies \citep[e.g.,][]{miyoshi1995, herrnstein2005, moran2008}. Such systems serve as powerful probes of the physical conditions influencing nuclear star formation. Being  luminous and intrinsically narrow, maser line emissions are a promising tool to measure the mass of the central supermassive black hole in these galaxies at the percent level \citep[e.g.,][]{gao2017}, as well as to provide a precise determination of the Hubble constant  from their kinematics, especially velocities and accelerations \citep{kuo2013, reid2013, humphreys2013}. 

Maser candidates are usually identified by their characteristic spectrum that displays distinct redshifted, systemic and blue-shifted components \citep{kuo2020}. The criterion for masing is the presence of a large velocity coherence along the line-of-sight. Some maser spectra show the presence of a double peak instead of a triple peak, whose physical nature is still not completely understood. In some double-peaked systems such as Mrk 1, the red- and blue-shifted masers show very different kinematics, with the kinematics of the redshifted masers being far more complicated and inconsistent with Keplerian rotation, and their speeds being systematically lower than those of the blueshifted masers. Both these properties are difficult to explain by simple models. \footnote{The above conclusion is, however, dependent on the choice for the dynamical centre of the system being located at the unweighted average position of all the maser spots. If the dynamical centre is assumed to be offset from this location, a Keplerian rotation curve may be consistent with both the blueshifted and redshifted masers, albeit with a different value of the best-fitting central black hole mass.} The profiles themselves are could potentially be attributed to triple-peaked systems with a weak systemic component \citep{kuo2011}. { Such maser distributions are likely to also be affected by non-gravitational forces such as Active Galactic Nuclei (AGN) winds or organized outflows. Examples include Mrk 1 \citep{kuo2020}, NGC 3079 which shows evidence for a parsec-scale  jet \citep{trotter1998}, and the hydrogen maser MWC 349 which shows complex velocity gradients perpendicular to the disk that are related to an ionized outflow \citep{martin2011}.} 

In this paper, we address the kinematics of double-peaked masers by comparing them to the properties expected for masing particles around a generic binary black hole system with black hole masses $5 \times 10^7$ and $5 \times 10^6 M_{\odot}$ respectively. With the help of simulations, we quantify the differences in the position-velocity distribution of masers around single and binary black holes, using the coherence length calibrated to the observations of NGC 4258 as the primary masing criterion. We find that both the velocity profiles as well as the line-of-sight velocity to impact parameter distributions in the binary black hole case show similarities to the observations of double-peaked masers such as Mrk 1. Our results motivate further searches for signatures of binary black hole candidates from the kinematics of water maser systems.

\section{Single black hole -  distribution maser sources}

One of the most well-studied examples of water masers in the literature is the galaxy NGC 4258 \citep{miyoshi1995, reid2019}. In what follows, we simulate masers sourced around a single black hole and calibrate the masing criterion from the observations of NGC 4258 \citep{argon2007}. 

We use the open-source N-body integrator software {\sc rebound}\footnote{https://rebound.readthedocs.io/en/latest/} \citep{rein2012} to track the positions, line-of-sight velocities (in the BH frame of reference), impact parameters and accelerations of $10^4$ clumps under the gravitational influence of a black hole of mass $M_{\rm bh} = 5 \times 10^7 M_{\odot}$. { The masses of the clumps are assumed to be small enough that their mutual gravitational attraction can be neglected.}

The simulation is initialized with the particles uniformly distributed between 0.3 pc and 6 pc from the central black hole, with their true anomalies randomly picked to lie between 0 and 2$\pi$. The disk is assumed to be edge-on and thin (corresponding to $h/r \leq 0.1$ as observed for the clumps in NGC 4258) and thus the inclinations of the particles are chosen to be randomly situated between 0 and 0.1 radians. The simulation is run { with a timestep  equal to one-tenth the orbital period of the nearest particles, and for a total time corresponding to 400 orbits of the farthest particles, which is found to lead to convergence of the distribution.}

\begin{figure}
\begin{center}
\includegraphics[width = \columnwidth]{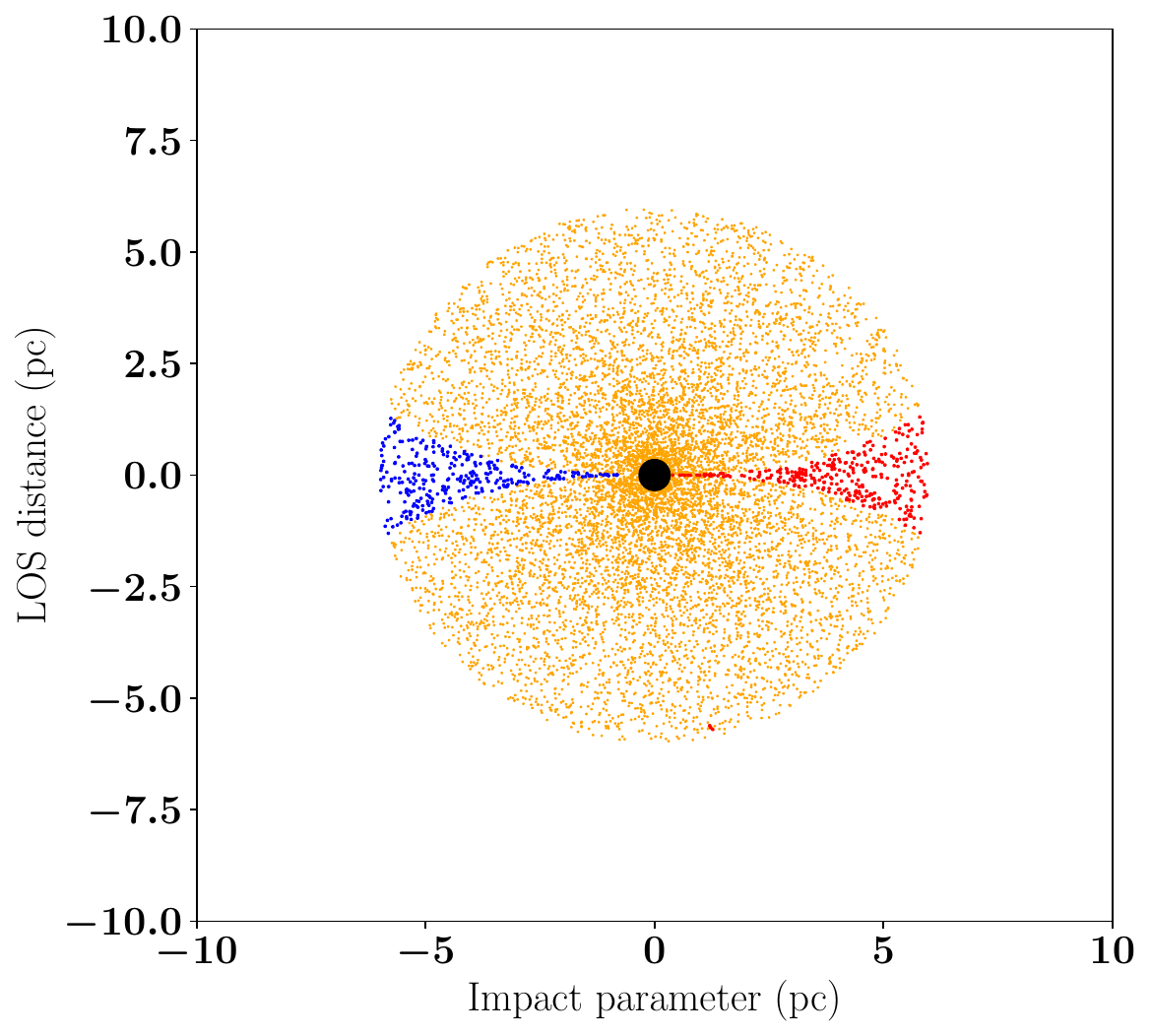}
\end{center}
\caption{Positions of simulated clumps (orange dots) around a single black hole (central black circle) of mass $5 \times 10^7 M_{\odot}$. The red and blue masing clumps on either side are indicated by their respective colors (following the definition in Sec. \ref{sec:binary} of the main text).}
\label{fig:1bhpositions}
\end{figure}

\begin{figure}
\begin{center}
  \includegraphics[width = 0.98\columnwidth]{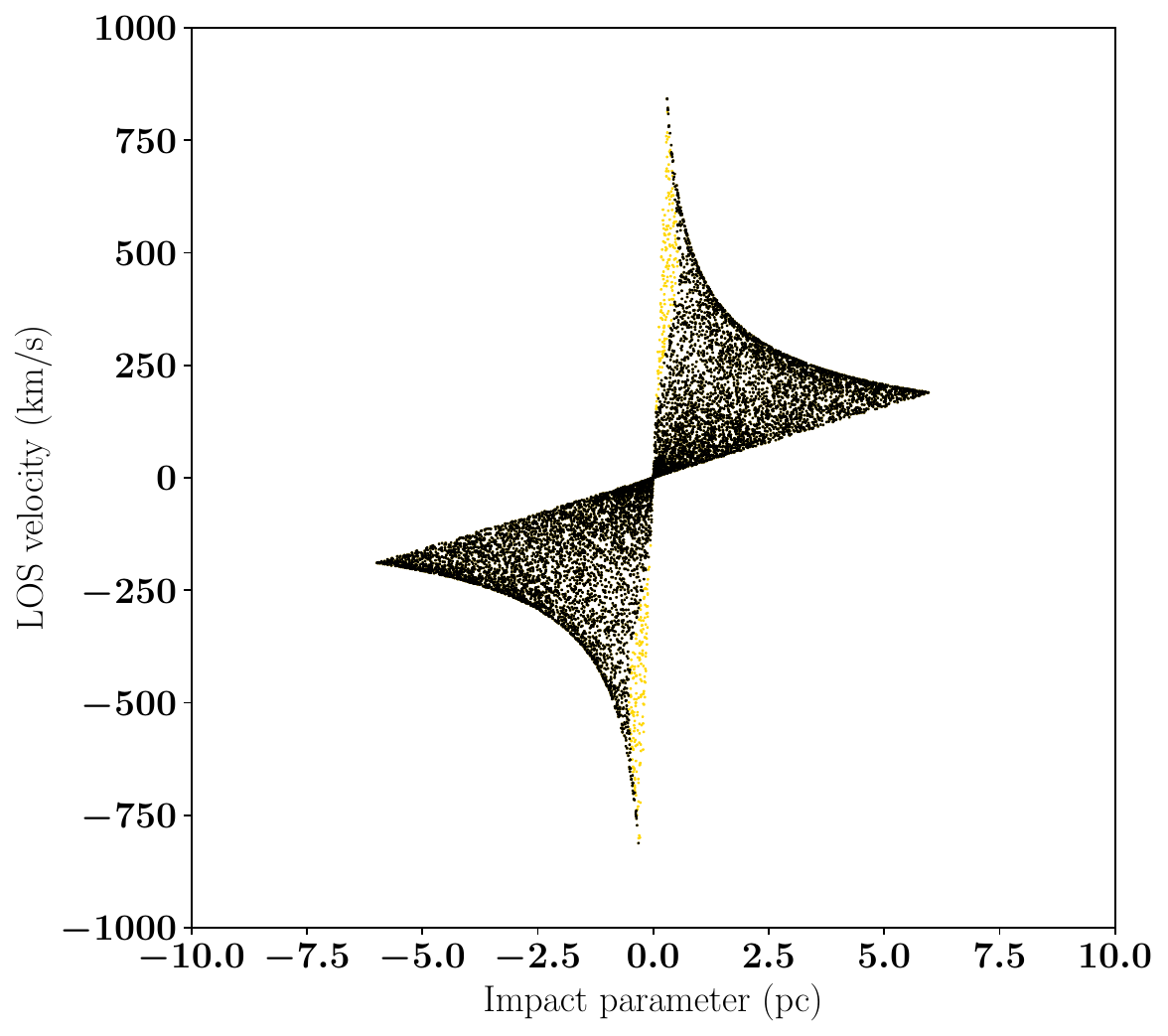} \includegraphics[width = 0.98\columnwidth]{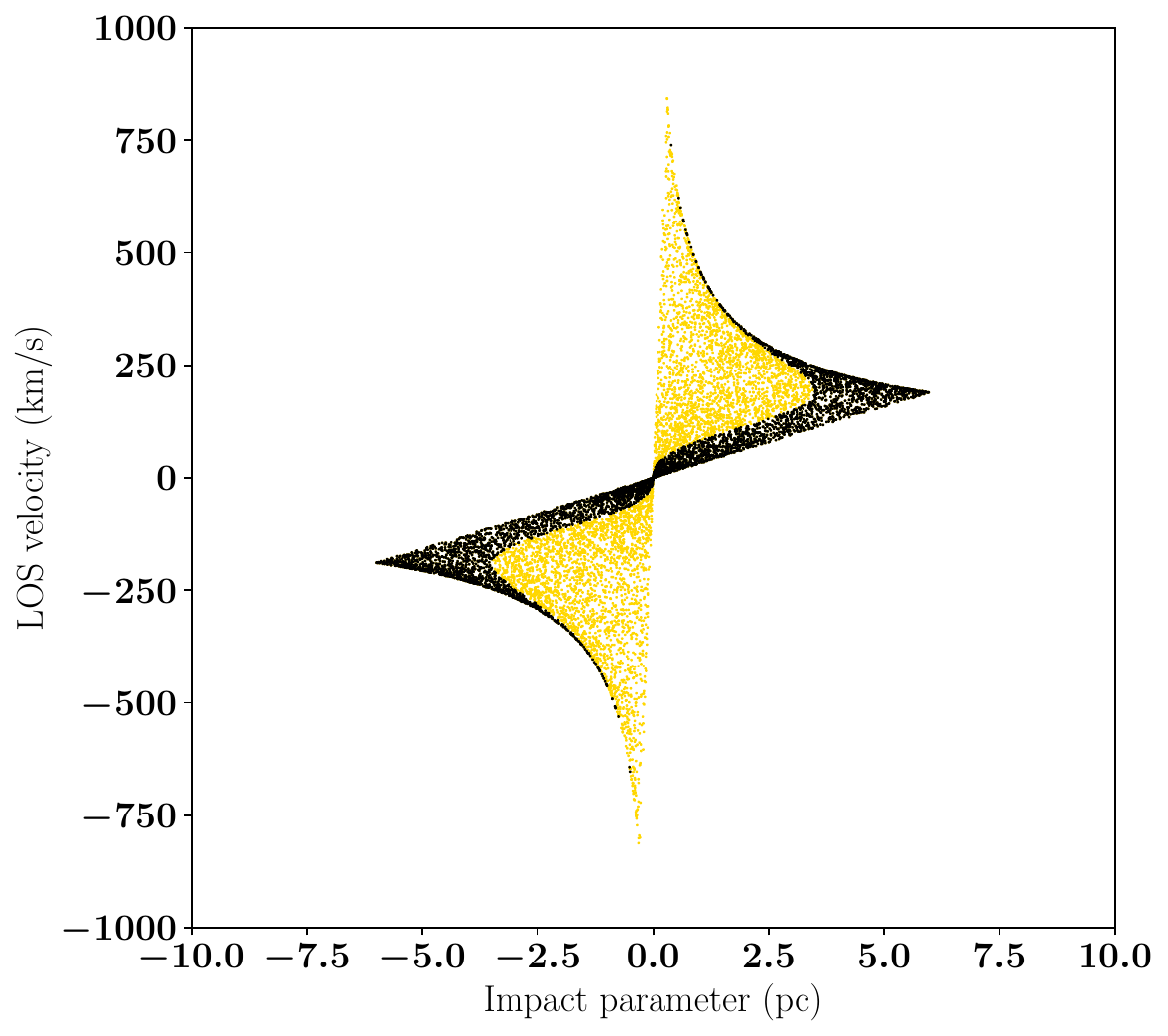} \includegraphics[width = 0.98\columnwidth]{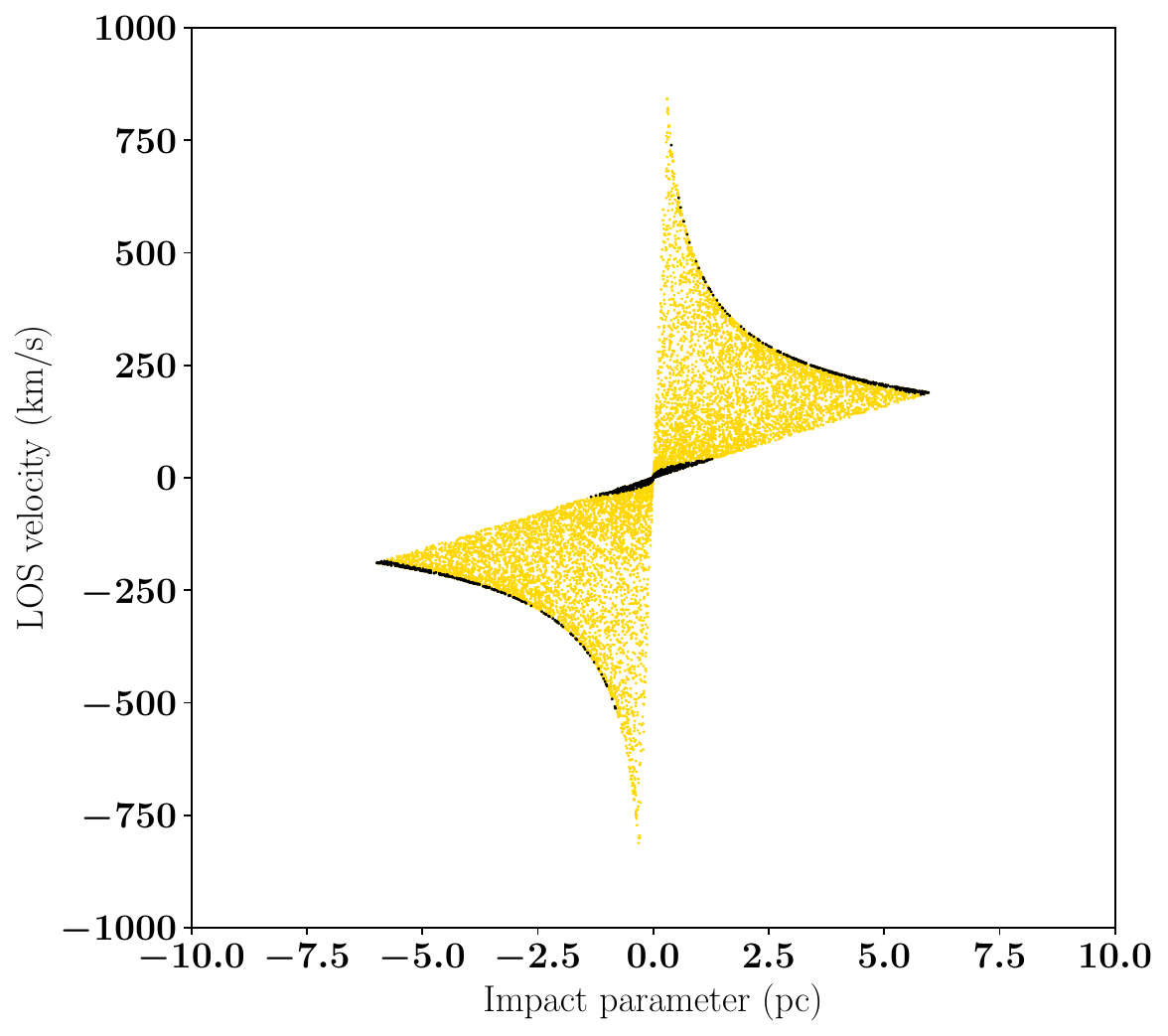}
 \end{center}
 \caption{ {{ Line-of-sight velocity ($v_{\rm LOS}$, in km/s) as a function of impact parameter ($b$, in pc) for the masers in the simulation. From top to bottom, the black points indicate the locations of the masers whose coherence length is greater than the 50th, 90th and 95th percentile respectively of that in NGC 4258.}}}
\label{fig:coherence4258}
\end{figure}

The positions of the clumps around the black hole at the end of the time period are plotted in Fig. \ref{fig:1bhpositions}. The impact parameter (defined as the projected distance to the masing particle along the plane of the disk) is plotted with respect to the line-of-sight velocity of all the simulated clumps as the gold points in Fig. \ref{fig:coherence4258}. 

Given the line-of-sight velocities (denoted by $v_{\rm LOS}$), we can express them in terms of the impact parameter $b$ as \citep{moran2008, humphreys2013}:
\begin{equation}
v_{\rm LOS} = \sqrt{\frac{GM_{\rm bh}}{R}} \sin i_r \cos \phi \approx \sqrt{\frac{GM_{\rm bh}}{R}} \frac{b}{R}
\label{vlos}
\end{equation}
where $\cos \phi = b/R$, $i_r$ denotes the { inclination} angle of the disk and $R$ is the distance of the clump to the central black hole.
In the last expression, we have approximated the system as being observed edge on, so that $i_r \approx 90^{\circ}$. In our simulations, we also assume circular orbits so that the eccentricity $e = 0$.

We can calculate the distribution of the velocity gradient ($dv_{\rm LOS}/dz$) along the line of sight as a function of the impact parameter. If the $z$-axis is along the line-of-sight, then 
\begin{equation}
R=\sqrt{b^2+z^2}
\end{equation}
where $b$ measures the separation of the line-of-sight from the black hole at closest approach, with the point of closest approach labelled by $z=0$. We have:

\begin{equation}
v_{\rm LOS} = \sqrt{GM_{\rm bh} /R} (b/R) 
\label{vlostobR}
\end{equation}
The line-of-sight of each clump is defined by $b= $ constant (since the observer is at a very large distance and the light rays are parallel near the source). Therefore, along each line-of-sight, $db=0$, and $2RdR = 2zdz$, 
and thus  $dz=(R/z)dR$. We then have:
\begin{eqnarray}
dv_{\rm LOS}/dz &=& (z/R)(dv_{\rm LOS}/dR)\nonumber \\
&=& (\sqrt{R^2-b^2}/R)(dv_{\rm LOS}/dR) \nonumber \\
&=& -1.5b (\sqrt{R^2-b^2}) \sqrt{GM_{\rm bh}} R^{-7/2} 
\label{vlosgradient}
\end{eqnarray}

From the velocity gradient we define the coherence of the clumps as:
\begin{equation}
L = c/|dv_{\rm LOS}/dz|
\end{equation}
This quantity is directly proportional to the optical depth $\tau$ of the masing systems [which is related to the observed flux{\footnote {{The exponential dependence holds for unsaturated masers. The relation is more complicated for saturated masers \citep[e.g.,][]{alcock1985, elitzur1996}, but this distinction is not important for our present purpose since only the strongest masing systems are considered for the analysis.}}} by $F = \exp(\tau)$], and hence its value serves as a criterion to determine the particles that are able to mase.

{ Several insights related to the above equations in the context of Fig. \ref{fig:coherence4258} are worth mentioning. From \eq{vlos}, we see that $v_{\rm LOS} \propto b/R^{1.5}$.
This dependence is shown in Fig. \ref{fig:coherence4258} by the straight lines demarcating the inner and outer radii of the masing disk. 
We also see from \eq{vlosgradient} that the two lines-of sight where the velocity gradient is zero correspond to $b=0$ and $b=R$. While the former shows up as the linear region at the centre of the plot, the latter corresponds to the black highlighted points situated along the curved, Keplerian boundary in the figures. These high-velocity regions are located at the `midline' of the disk, at which $v_{\rm LOS} = (GM/R)^{1/2}$.}

\begin{figure}
 \begin{center}
  \includegraphics[width = \columnwidth]{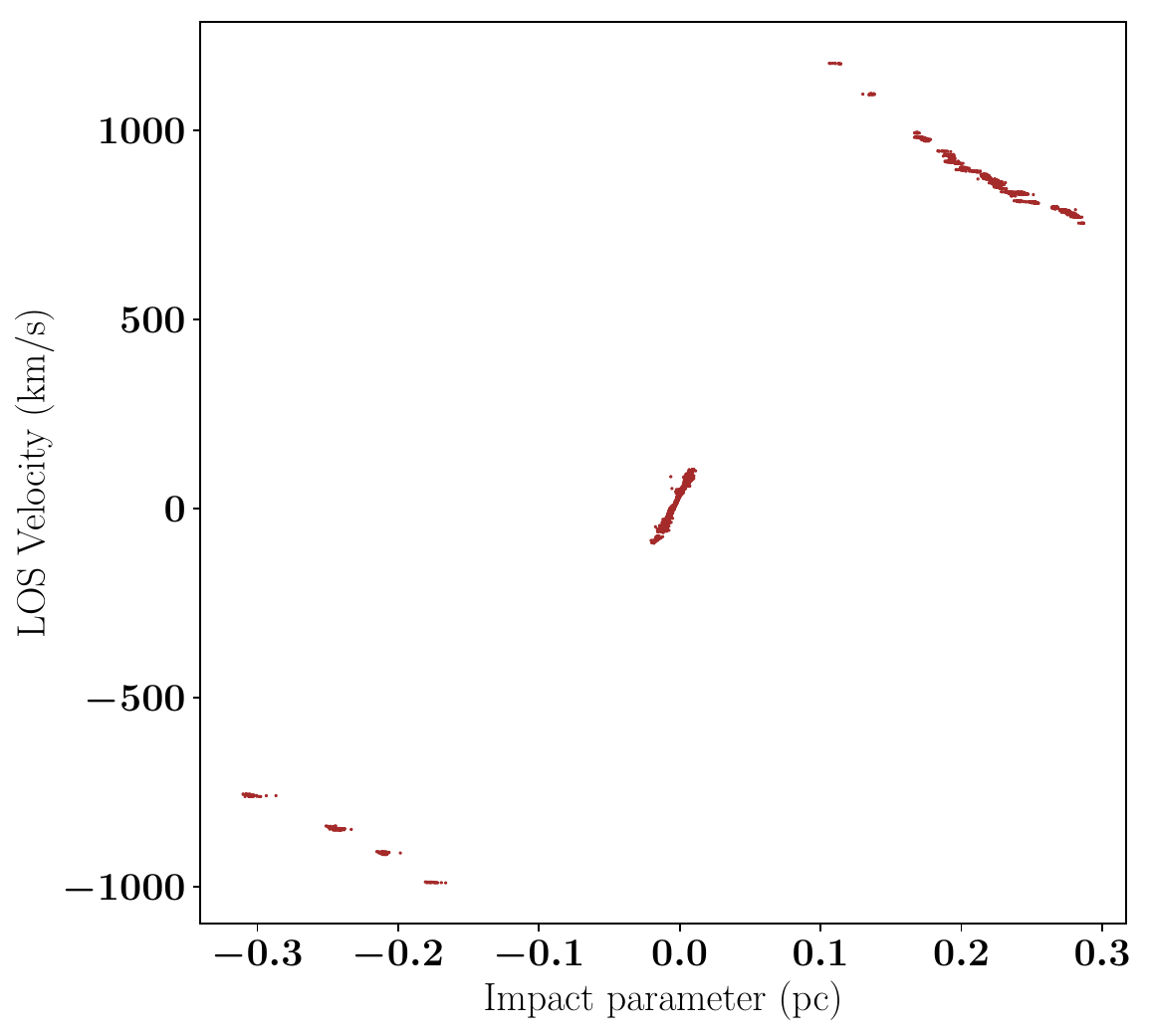} 
 \end{center}
 \caption{Line-of-sight velocities as a function of impact parameter for the masing clumps in NGC 4258. The velocities are in the black hole's  frame of reference, with the data taken from \citet{argon2007}.}
\label{fig:impactlosngc4258}
\end{figure}

We now calibrate the coherence criterion by comparing to the observations of a known water maser galaxy, NGC 4258 \citep{herrnstein2005}. The velocities and impact parameters along the line of sight as given in the observations of \citet{argon2007} are plotted in Fig. \ref{fig:impactlosngc4258}. { The single straight line in the center of the figure is attributable to the disk in NGC 4258 being warped \citep{herrnstein2005, martin2008} leading to a `bowl'  shape in the direction of the observer \citep{moran2008} that confines the features to a thin annular region.} The velocities are plotted in the  black hole frame of reference, i.e. 
\begin{equation}
v_{\rm LOS} = v_{\rm rel} - v_0
\end{equation}
where $v_{\rm rel}$ is the observed velocity and $v_0 = 474.25$ km/s is the observed systemic velocity of NGC 4258.  
We now use \eq{vlostobR} to find the value of $R$ of each clump (based on its measured values of $b$ and $v_{\rm LOS}$), which can be substituted into the last expression in deriving the value of $dv_{\rm LOS}/dz$ of that clump, and thus the observed coherence distribution.
The particles in the simulation that have coherences greater than the 50th, 90th and 95th percentile respectively of the coherence distribution in NGC4258 are shown by the black points  in the top, middle and lower panels of Fig. \ref{fig:coherence4258}. It is seen that the particles having coherences greater than the 95th percentile are visually consistent with the observed maser distribution.

\section{Black hole binaries}
\label{sec:binary}

Having calibrated the coherence criterion defining the masing particles to the observations of NGC 4258, we can now use it to identify their analogues  around a binary black hole system. To do this, we use {\sc rebound} to simulate $10^4$ masing clumps around a system of 2 black holes, with the black hole masses in the ratio $q:1$ (with $q$ being the mass ratio that can take different values between 0 and 1) with the more massive black hole having $M_{\rm bh} = 5 \times 10^7 M_{\odot}$, and separated by $a = 3$ pc. \footnote{We consider separations of the parsec scale here, since a large population of supermassive black hole binaries are expected to be found in this range [the `final parsec' problem, \citet{begelman1980, milosavljevic2003}]. For the masses of the black holes under consideration, $\sim 1$ pc also represents the `hardening' scale of the binary \citep[e.g.,][]{yu2002}. The scale at which gravitational wave emission takes over is much smaller, of the order of a few milliparsec for the masses considered here \citep{peters1964, loeb2010}. Hence, these binaries are expected to spend most of their time around the parsec radius.} 

\begin{figure}
\begin{center}
\includegraphics[width = \columnwidth]{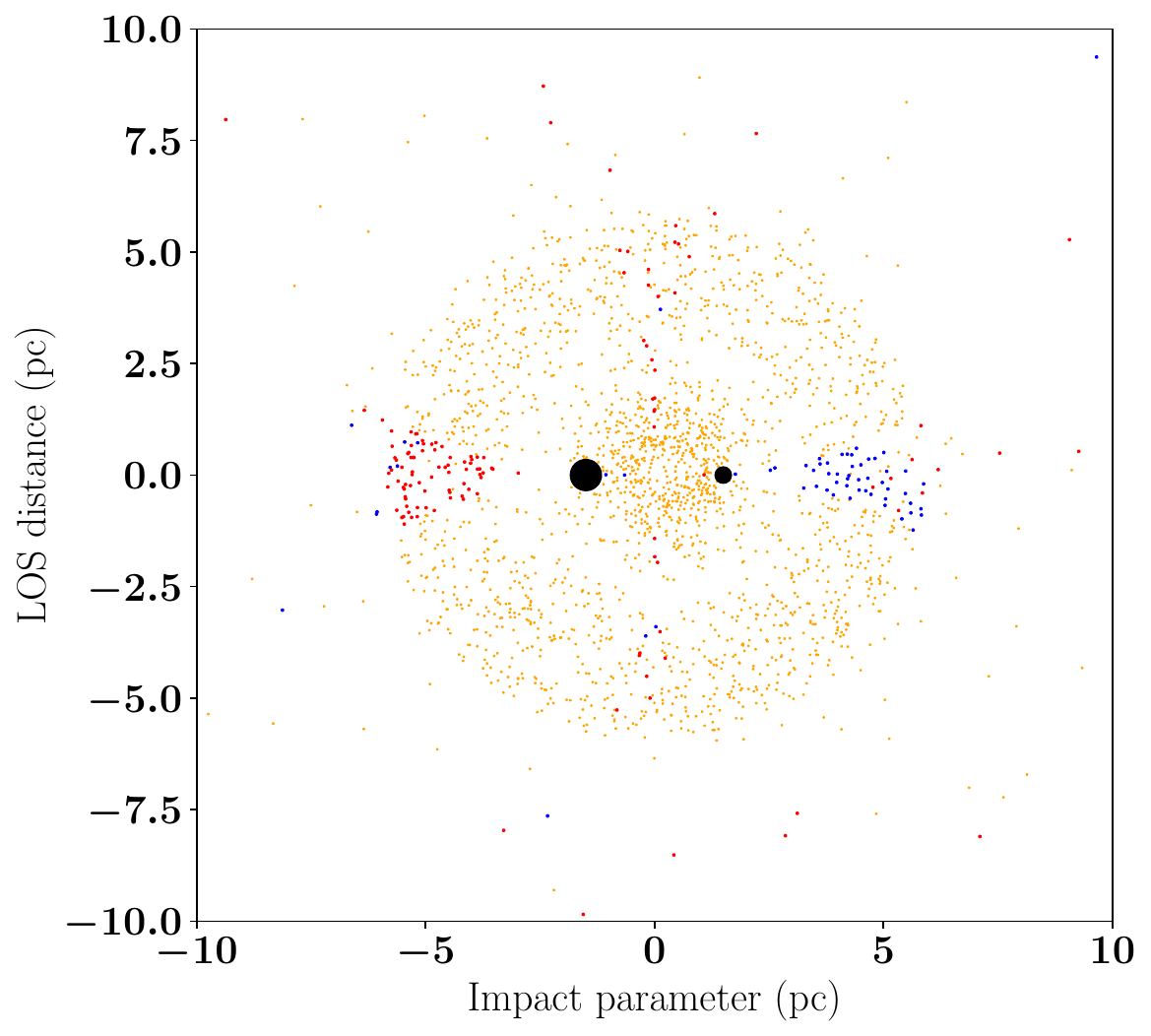}
\end{center}
\caption{Positions of simulated clumps (orange dots) around the black hole binary having masses $5 \times 10^7 M_{\odot}$ and $5 \times 10^6 M_{\odot}$ (black circles) separated by $a = 3$ pc. The red and blue masing clumps are indicated by their respective colors (following the definition in the main text).}
\label{fig:2bhpositions}
\end{figure}

\begin{figure}
 \begin{center}
  \includegraphics[width = \columnwidth]{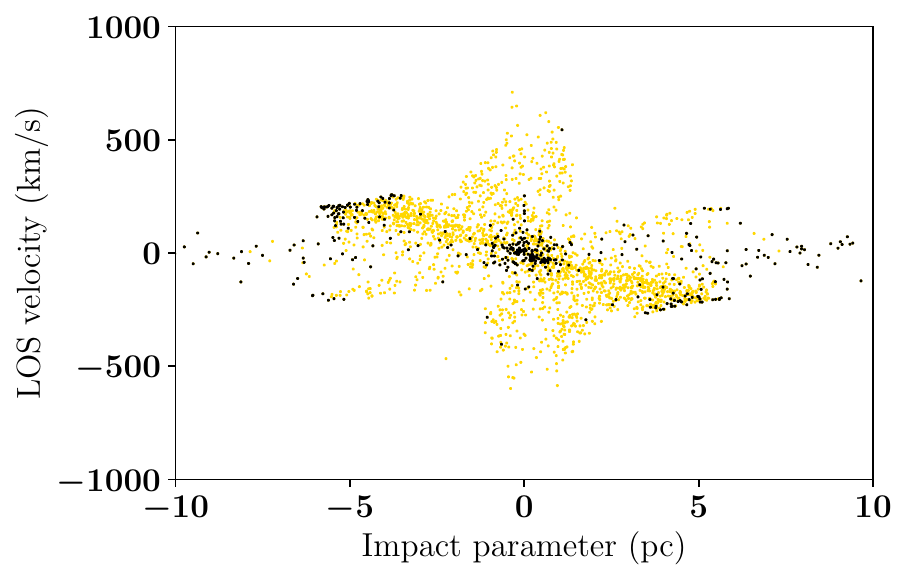} 
 \end{center}
 \caption{Line-of-sight velocity as a function of impact parameter for simulated clumps around a binary black hole with masses $5 \times 10^7 M_{\odot}$ and $5 \times 10^6 M_{\odot}$ separated by $a = 3$ pc. Particles whose coherence lengths are greater than the 95th percentile of the observations of NGC 4258 are marked in black.}
\label{fig:masingclumps2bh}
\end{figure}

We describe the case $q = 0.1$ in the analyses below. For this case, the second black hole has mass $5 \times 10^6 M_{\odot}$. The positions of the clumps around the black hole binary are plotted in Fig. \ref{fig:2bhpositions}. The impact parameter is plotted with respect to the line-of-sight velocities of all the simulated clumps as the { gold} points in Fig. \ref{fig:masingclumps2bh}. 

The number of masers which are expected around such a system is now found by restricting to only those values of coherence which are greater than the 95th percentile of the coherence distribution of NGC 4258. The corresponding maser locations are plotted as the { black} points in in Fig. \ref{fig:masingclumps2bh}. The figure shows that the position - velocity distribution shows distinct differences from that of a single black hole shown in the lower panel of Fig. \ref{fig:coherence4258}. This serves as a potential signature to distinguish the presence of a binary black hole if the masing clumps can be resolved.  

\begin{figure}
\begin{center}
\includegraphics[width = \columnwidth]{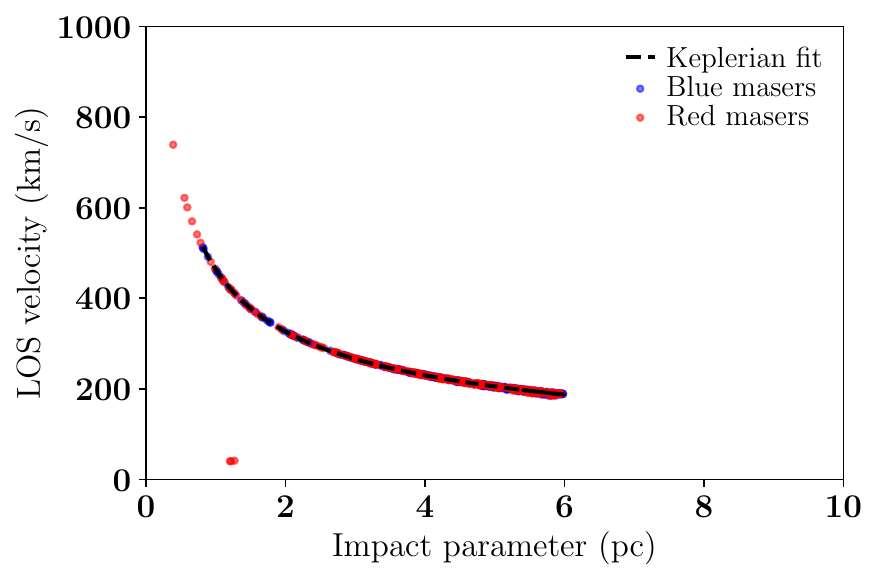}
\end{center}
\caption{Absolute values of line-of-sight velocity as a function of impact parameter for the simulated red and blue masing particles (defined as those having $v_{\rm LOS} > 40$ km/s and $v_{\rm LOS} < -120$ km/s respectively) around a single black hole of mass $5 \times 10^7 M_{\odot}$. The dashed black line shows the best-fitting Keplerian rotation curve which is consistent with both groups of particles.}
\label{fig:1bhkepler}
\end{figure}

It was recently found that \citep{kuo2020} the galaxy Mrk 1 exhibits a double peaked structure of masing particles.  Depending on the choice of the gravitational centre, the blue-shifted masers may be consistent with a Keplerian curve, with the red-shifted ones showing 
evidence of an offset and also being scattered around lower values of line-of-sight velocities. 

\begin{figure}
\begin{center}
\includegraphics[width = \columnwidth]{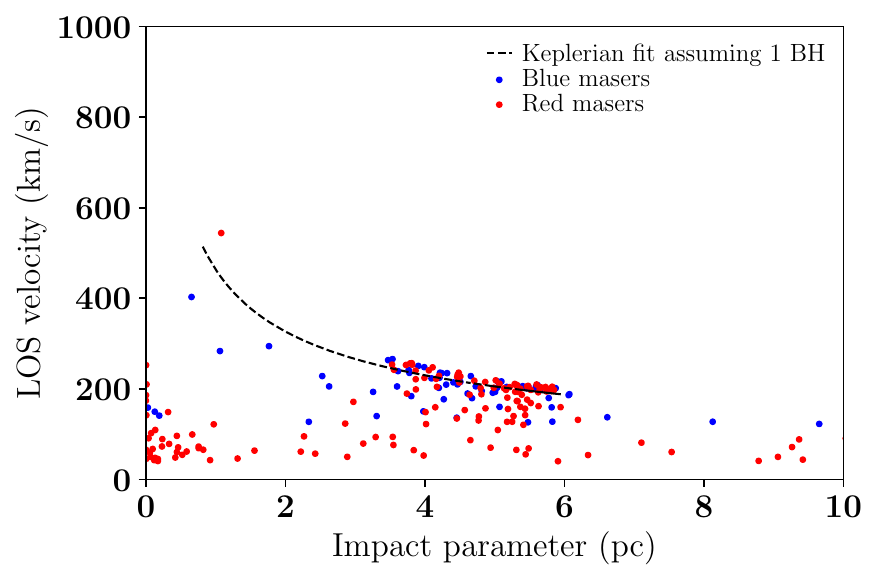}
\end{center}
\caption{Absolute values of line-of-sight velocity as a function of impact parameter for simulated red and blue masing particles (defined as those having $v_{\rm LOS} > 40$ km/s and $v_{\rm LOS} < -120$ km/s respectively) around a binary black hole system with masses $5 \times 10^7 M_{\odot}$ and $5 \times 10^6 M_{\odot}$. The dashed black line shows the best-fitting Keplerian rotation curve fitted to all the masing particles in the simulations of a single black hole in Fig. \ref{fig:1bhkepler}.}
\label{fig:2bhkepler}
\end{figure}

To explore this effect in the simulations, we split the masing particles into red and blue subsets to match the criterion adopted in the observations of Mrk 1, i.e. $v_{\rm LOS} > 40$ km/s and $v_{\rm LOS} < -120$ km/s respectively. The absolute value of the line-of-sight velocity is shown as a function of the impact parameter for both these sets of particles in Figs. \ref{fig:1bhkepler}  and \ref{fig:2bhkepler}, for the single and binary black hole cases respectively. The dashed black line shows the best-fitting Keplerian fit to all the masing particles around the single black hole. 

It is seen that while the blueshifted masers follow the Keplerian rotation curve in both cases, the redshifted ones deviate, notably with a scatter around low line-of-sight velocities, in the binary black hole case. This finding lends support to a binary black hole interpretation (with a mass ratio of $1:10$) to systems such as Mrk 1 whose kinematics show a similar trend for the red and blue clumps. 

\begin{figure}
\begin{center}
\includegraphics[width = \columnwidth]{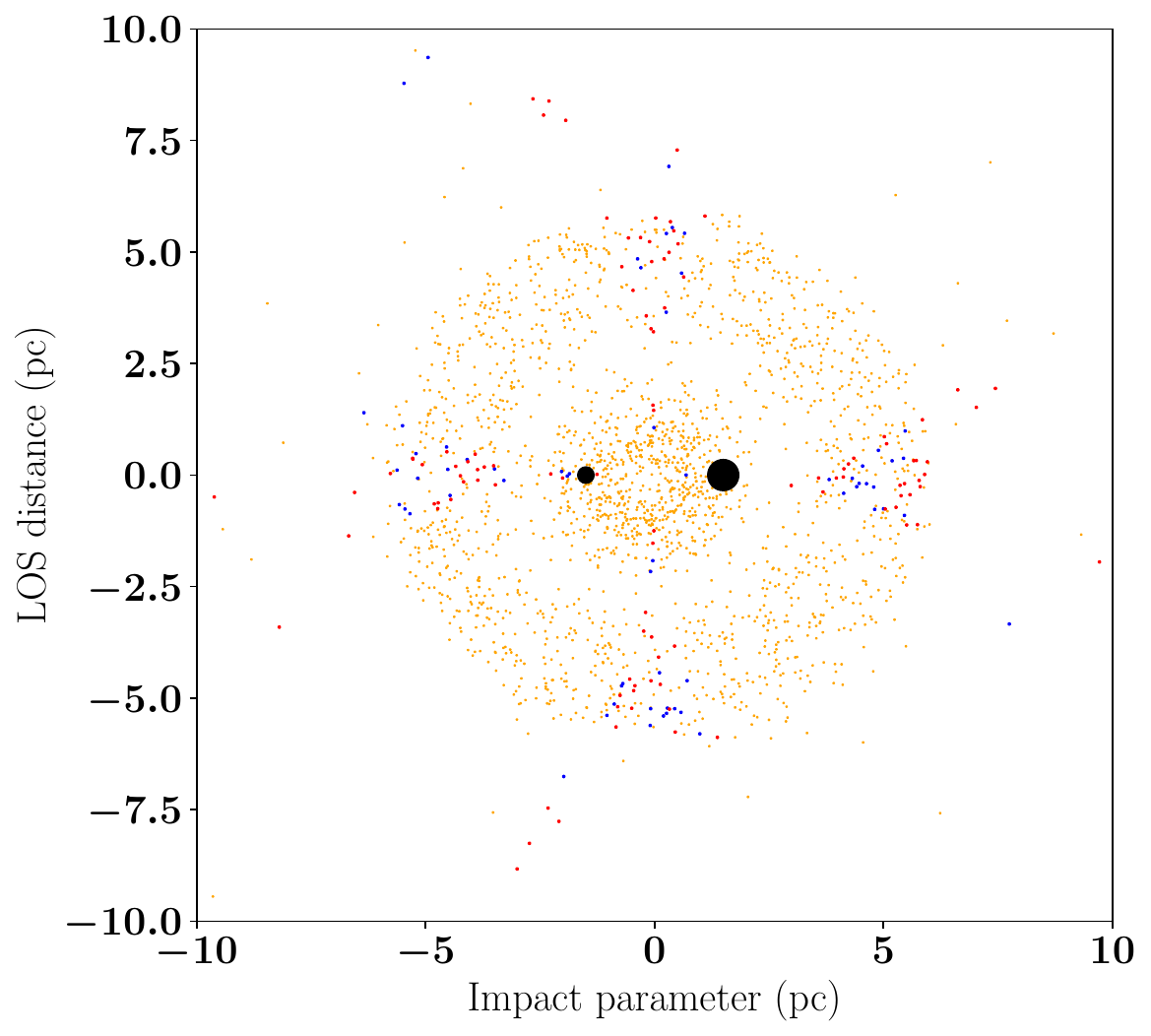}
\end{center}
\caption{Same as Fig. \ref{fig:2bhpositions}, but with the positions of the two black holes reversed.}
\label{fig:2bhpositionsreversed}
\end{figure}

\begin{figure}
\begin{center}
\includegraphics[width = \columnwidth]{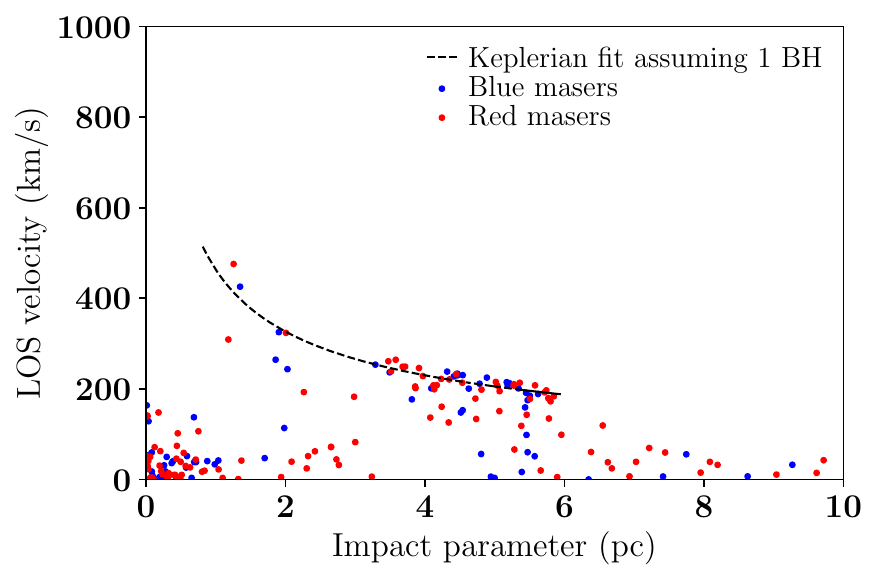}
\end{center}
\caption{Same as Fig. \ref{fig:2bhkepler}, but with the positions of the two black holes reversed.}
\label{fig:2bhkeplerreversed}
\end{figure}

It is to be noted, however, that the scatter observed in the impact parameter - velocity is a function of the orientation (or equivalently phase of orbit) apart from the criterion by which the colors are defined. To illustrate this, we plot in Fig. \ref{fig:2bhpositionsreversed} the locations of all the masing clumps and the two black holes in the opposite orientation, and the corresponding absolute line-of-sight velocity as a function of impact parameter in Fig. \ref{fig:2bhkeplerreversed}. The plots still show considerable scatter but now with more blue masing clumps deviating from the Keplerian prediction (with the definitions of red and blue clumps being identical to those used in Fig. \ref{fig:2bhkepler}.)

\begin{figure}
\begin{center}
\includegraphics[width = \columnwidth]{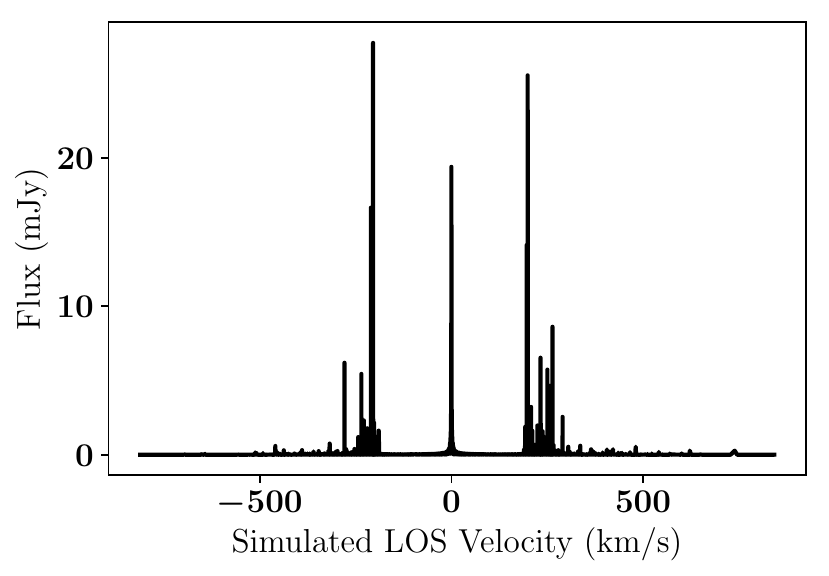}
\end{center}
\caption{Simulated line-of-sight spectra for the masing particles  around a central black hole of mass $5 \times 10^7 M_{\odot}$. The flux is computed by normalizing the simulated coherences of each particle to the optical depths observed for the masers around NGC 4248.}
\label{fig:velprofile1bh}
\end{figure}

\begin{figure}
\begin{center}
\includegraphics[width = \columnwidth]{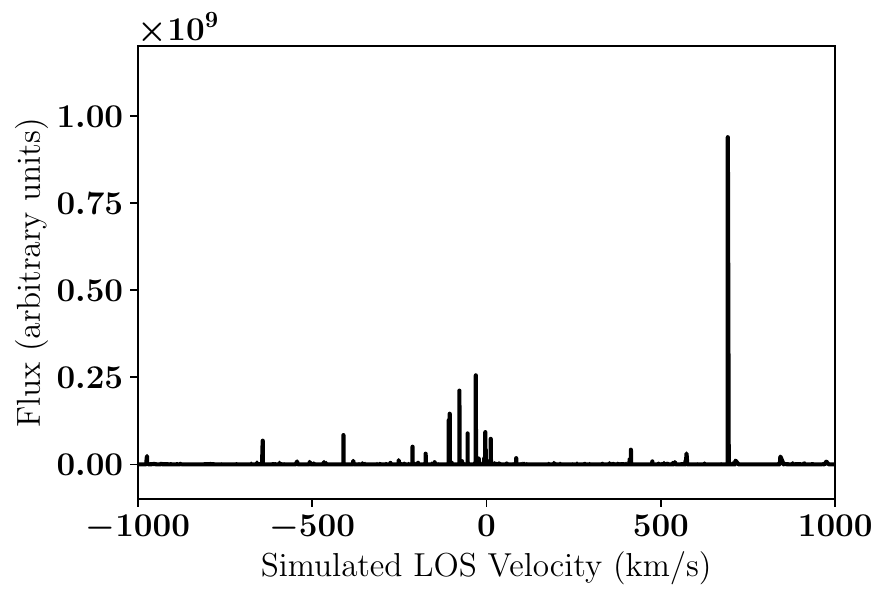}
\end{center}
\caption{Same as Fig. \ref{fig:velprofile1bh}, but for the masing particles  around a binary black hole system with masses $5 \times 10^7 M_{\odot}$ and $5 \times 10^6 M_{\odot}$  separated by $a = 3$ pc.}
\label{fig:velprofile2bh}
\end{figure}

We also model the line-of-sight spectra to illustrate their differences for the single and binary central black hole cases. To do this, we normalize the absolute value of the coherence at each masing location to the observed mean optical depth of NGC 4258, and use the normalized value to compute the resultant flux as a function of the simulated velocity along the line-of-sight. The resultant spectra for the two cases are shown in Figs. \ref{fig:velprofile1bh} and \ref{fig:velprofile2bh} respectively. Since the normalization in the binary black hole case is arbitrary, the units for the flux are arbitrary in this case. It can be seen that the binary black hole case is consistent with a double-peaked nature of the maser spectrum as reported in observations.This lends further support to the interpretation of such systems as surrounding a binary, rather than a single black hole.

\section{Conclusions}

Water masers are a unique probe of several physical processes taking place in active galaxies, several of which are not yet well understood { \citep[for a review, see e.g., ][]{lo2005}}. These systems are nevertheless important tools to infer the masses of the central supermassive black holes as well as to provide tests of cosmology by enabling a precise determination of quantities such as the Hubble constant.

In this paper, we have compared the dynamics of water maser systems in the environments of single and binary supermassive black holes.   { We have illustrated how binary black
hole systems present themselves in maser maps and position-velocity diagrams, facilitating the use of maser emission as a tracer of their dynamics.
 } By calibrating the simulated kinematics of masers around a single black hole to the observations of NGC 4258, we established the criterion for clumps to mase based on their coherence length being greater than a given percentile of the observed distribution. We used this to search for evidence of the presence of binary black holes in the class of double-peaked maser systems, finding similarities between the simulations and observations in both the kinematics and the spectra. Our results suggest that the physical nature of double peaked and similar masing systems could be explained by the perturbing influence of a second black hole.{\footnote{ Note that double-peaked structures such as in Fig. \ref{fig:velprofile2bh} with the cluster of particles close to $v_{\rm LOS} = 0$ could also be interpreted as missing the blue component in a non-binary black hole system.}}

In this work, we considered a simple illustrative example for pedagogical purposes. More generally, a binary companion may orbit on a plane different from the masing clumps. This could potentially trigger the observed warp in the outer part of NGC 4258 for a perturber outside the masing disk \citep{papaloizou1998}.

The water masers can potentially act as electromagnetic counterparts for gravitational waves from their supermassive black hole mergers to localize the central galaxy. Note, however that the characteristic strain associated with the merger of black holes  -- with masses and separations in the range we consider here -- is $h_0 \sim 10^{-20}$, almost four orders of magnitude lower than the currently measured upper limits from Pulsar Timing Arrays \citep{verbiest2021}. The frequencies are also 2-4 orders of magnitude lower than the corresponding ones at which present upper limits exist (10 nHz and 31 nHz respectively). For these reasons, the kinematics of these systems acts as the unique probe of the presence of a binary black hole in this regime of masses and separations.
High-resolution imaging of the central regions in such systems \citep[e.g.,][]{kamali2019} would serve as a useful follow-up to confirm the presence of a binary companion.

\section*{Acknowledgements}  

We thank Mark Reid for several useful discussions at various stages of this work and for comments on the manuscript, Hanno Rein for for helpful clarifications related to {\sc rebound} { and the referee for a detailed and constructive report that improved the content and presentation}. HP acknowledges support from the Swiss National Science Foundation through Ambizione Grant PZ00P2\_179934. The work of AL is partially supported by Harvard's Black Hole Initiative, which is funded by grants from JTF and GBMF.

\section*{Data availability}
No new data were generated or analysed in support of this research. The software underlying this article will be shared on reasonable request to the authors.

\def\aj{AJ}                   
\def\araa{ARA\&A}             
\def\apj{ApJ}                 
\def\apjl{ApJ}                
\def\apjs{ApJS}               
\def\ao{Appl.Optics}          
\def\apss{Ap\&SS}             
\def\aap{A\&A}                
\def\aapr{A\&A~Rev.}          
\def\aaps{A\&AS}              
\def\azh{AZh}                 
\def\baas{BAAS}
\def\jcap{JCAP}
\def\jrasc{JRASC}             
\def\memras{MmRAS}
\def\na{New Astronomy}
\def\nat{Nature}
\def\mnras{MNRAS}             
\def\pra{Phys.Rev.A}          
\def\prb{Phys.Rev.B}          
\def\prc{Phys.Rev.C}          
\def\prd{Phys.Rev.D}          
\def\prl{Phys.Rev.Lett}       
\def\pasp{PASP}               
\def\pasj{PASJ}
\def\physrep{Phys. Repts.}
\def\qjras{QJRAS}             
\def\skytel{S\&T}             
\def\solphys{Solar~Phys.}     
\def\sovast{Soviet~Ast.}      
\def\ssr{Space~Sci.Rev.}      
\def\zap{ZAp}                 
\let\astap=\aap
\let\apjlett=\apjl
\let\apjsupp=\apjs

\bibliography{mybib}{}
\bibliographystyle{mnras}

\end{document}